\documentclass[
pra,aps,
reprint,
a4paper,
superscriptaddress,
longbibliography,
floatfix
]{revtex4-1}
\usepackage{times}
\usepackage{graphicx}
\usepackage{epsfig}
\usepackage{amsfonts}
\usepackage{amsmath}
\usepackage{amssymb}
\usepackage{color}
\usepackage{multirow}
\usepackage[colorlinks=true,linkcolor=blue,citecolor=magenta,urlcolor=blue]{hyperref}
\def\endproof{\vrule height6pt width6pt depth0pt}

\begin{document}


\title{Necessary and sufficient condition for contextuality from incompatibility}


\author{Zhen-Peng Xu}
\affiliation{Theoretical Physics Division, Chern Institute of Mathematics,
	Nankai University,
	Tianjin 300071, People's Republic of China}
\affiliation{Departamento de F\'{\i}sica Aplicada II,
	Universidad de Sevilla,
	E-41012 Sevilla, Spain}

\author{Ad\'an Cabello}
\email{adan@us.es}
\affiliation{Departamento de F\'{\i}sica Aplicada II, Universidad de Sevilla, E-41012 Sevilla, Spain}


\begin{abstract}
	Measurement incompatibility is the most basic resource that distinguishes quantum from classical physics. Contextuality is the critical resource behind the power of some models of quantum computation and is also a necessary ingredient for many applications in quantum information. A fundamental problem is thus identifying when incompatibility produces contextuality. Here, we show that, given a structure of incompatibility characterized by a graph in which nonadjacent vertices represent incompatible ideal measurements, the necessary and sufficient condition for the existence of a quantum realization producing contextuality is that this graph contains induced cycles of size larger than three.
\end{abstract}


\maketitle


{\em Incompatibility versus contextuality.} Measurement incompatibility is arguably the most basic resource that distinguishes quantum and classical physics. Incompatibility is ubiquitous in protocols with a quantum-over-classical advantage and has been proven to be necessary for no-cloning \cite{HMZ16} and nonlocality \cite{KC85,WPF09,QBHB16}. On the other hand, contextuality (a concept resulting from the Kochen-Specker theorem \cite{Specker60,Bell66,KS67}, but here used in the exact sense used in Refs.~\cite{KCBS08,Cabello08,YO12,KBLGC12,AQB13,CSW14}) is the critical resource behind the quantum advantage of some models of quantum computation \cite{AB09,HB11,Raussendorf13,HWVE14,DGBR15,ASM17,BDBOR17,RBDOB17} and a necessary ingredient for many quantum protocols (e.g., device-independent quantum key distribution \cite{Ekert91,BHK05}, quantum advantage in zero-error classical communication \cite{CLMW10}, and some cryptographic protocols \cite{CDNS11}). Therefore, a fundamental question is what is the relation between incompatibility and contextuality. This is the problem we address in this Rapid Communication

The definition of measurement incompatibility is independent of any physical theory. Two measurements, $A$, with outcome set $\{a_x\}_{x \in X}$, and $B$, with outcome set $\{b_y\}_{y \in Y}$, are incompatible (or not jointly measurable) if there is no measurement $C$ with outcome set $\{c_{x,y}\}_{x \in X, y \in Y}$ such that, for all initial states $\rho$, the probability $P(a_x|\rho)= \sum_{y \in Y} P(c_{x,y}|\rho)$, for all outcomes $a_x$, and the probability $P(b_y|\rho)=\sum_{x \in X} P(c_{x,y}|\rho)$, for all outcomes $b_y$. If such a $C$ exists, then $A$ and $b$ are compatible (or jointly measurable). In other words, two measurements $A$ and $B$ are incompatible if there does not exist a measurement $C$ such that both $A$ and $B$ are coarse grainings of $C$.

A measurement scenario is characterized by a set ${\cal M}$ of measurements, their respective outcomes, and the subsets of ${\cal M}$ that are compatible. The relations of compatibility between the measurements in a scenario are usually represented by a hypergraph in which each vertex represents a measurement and vertices in the same hyperedge are mutually compatible (see, e.g., Refs.~\cite{HRS08,HFR14,AT18}).

In general, contextuality indicates that the outcome statistics of an experiment involving several contexts (i.e., sets of compatible measurements) cannot be explained assuming that the outcomes reveal preexisting values that are independent of the context. However, there are several definitions of contextuality in the literature. The one for which a crucial connection with quantum computation has been established \cite{HWVE14} is the one used in Refs.~\cite{KCBS08,Cabello08,YO12,KBLGC12,AQB13,CSW14}. Given a measurement scenario where all measurements are ideal, a behavior (i.e., a set of probability distributions, one for each context) is contextual if it does not belong to the polytope whose vertices are all possible deterministic assignments of outcomes to the measurements in that scenario. A measurement is ideal (or sharp) \cite{CY14,CY16} if
(i) it yields the same outcome when performed consecutive times,
(ii) it only disturbs measurements that are incompatible with it, and
(iii) all its coarse grainings have realizations satisfying (i) and (ii).
In quantum theory, an ideal measurements is represented by a self-adjoint operator $A$ on a Hilbert space or, equivalently, by the set of orthogonal projectors
(onto distinct, possibly degenerate, eigenspaces of $A$) summing to the identity in the spectral decomposition of $A$. On the other hand, compatible measurements are represented in quantum theory by commuting operators.

The restriction of the definition of contextuality to scenarios involving only ideal measurements obeys three main reasons: (I)~It assures that compatible measurements do not disturb each other (which is what naturally happens in Bell scenarios due to the fact that measurements are spatially separated). A measurement $A$ disturbs a measurement $B$ if, for some initial state, from the outcome statistics of $B$, one can detect whether $A$ was performed. Recall that, for nonideal measurements, compatibility does not imply nondisturbance \cite{HRS08,HW10,HFR14}. (II)~It assures that the contextuality of a behavior can be taken as a signature of nonclassicality. On the one hand, as pointed out in Ref.~\cite{Spekkens14}, the assumption that the outcome of a measurement depends deterministically on the ontic state (which is the assumption satisfied by the extreme points of the set of noncontextual behaviors) is reasonable if and only if the measurement is ideal. In particular, it is not a physically plausible assumption when applied to a noisy measurement (even a classical one), since, in this case, the outcome may have an indeterministic dependence on the ontic state of the measured system. On the other hand, the classical simulation of quantum contextuality for ideal measurements has a quantifiable memory~\cite{KGPLC11,CGGX18} and thermodynamical overcosts~\cite{CGGLK16}. (III)~It assures that the classical and quantum sets of behaviors are direct generalizations of the corresponding sets for Bell scenarios. In particular, for contextuality scenarios that, by spacelike separating the measurements, can be converted into Bell scenarios, the sets of behaviors are identical regardless of whether there is timelike or spacelike separation.

For ideal measurements, if in a set of measurements every two of them are compatible, then all of them are compatible \cite{CY14} (this is not true for nonideal measurements \cite{Specker60,HRS08}). As a consequence, the relations of compatibility between ideal measurements can be represented by a simple graph, called compatibility graph, in which any clique of vertices represents a set of compatible measurements (see, e.g., Refs.~\cite{AT18,KRK12,CDLP13}). A clique of a graph is a set of vertices every pair of which are adjacent.

The fundamental problem is what is the relation between incompatibility and contextuality. Clearly, incompatibility is necessary for contextuality. Otherwise, if all measurements are compatible, then there is only one context. However, not every set of measurements that includes incompatible measurements produces contextuality. Therefore, the crucial question is what incompatibility structures can produce quantum contextuality and which ones cannot. Surprisingly, we have not found the answer to this question in the literature.

A first step towards solving this problem is a theorem introduced by Vorob'yev \cite{Vorob'yev63,Vorob'yev67} that has been used in connection to quantum theory in Refs.~\cite{BM10,BC12,RSKK12,Barbosa14,Barbosa15,BMC16}. The theorem states that, for any set of measurements whose corresponding compatibility graph is chordal (i.e., does not contain induced cycles of size larger than three), there is always a joint probability distribution for every behavior (see the Appendix). Therefore, in this case, all quantum behaviors can be simulated by a noncontextual hidden variable model. Recall that an induced subgraph of a graph $G(V,E)$, with vertex set $V$ and edge set $E$, is a graph with vertex set $S \subseteq V$ and edge set comprising all the edges of $G$ with both ends in $S$. An $n$-vertex cycle, denoted $C_n$, is a graph with $n$ vertices connected in a closed chain, e.g., $C_4$ is a square and $C_5$ is a pentagon. Therefore, a necessary condition for quantum contextuality is that the compatibility graph is not chordal.


{\em Main result.} The aim of this Rapid Communication is to prove and explore the consequences of the following result.


{\em Theorem.}
For a given compatibility graph $G(V,E)$, with vertex set $V$ and edge set $E$, there is a set of quantum ideal measurements $\mathcal{M} = \{M_i\}_{i\in V}$ satisfying the incompatibility/compatibility structure given by $G(V,E)$ and producing contextuality if and only if $G(V,E)$ is not a chordal graph.


{\em Proof.} That the nonchordality of the compatibility graph is a necessary condition for contextuality follows from the proof of Vorob'yev's theorem (see the Appendix). That nonchordality of the compatibility graph is a sufficient condition for contextuality can be proven as follows. Let $G_{1}(V_1, E_1)$ be the compatibility graph of ${\cal M}_1 = \{M_i\}_{i \in V_1}$. Let $G_{2}(V_2, E_2)$ be a compatibility graph such that $G_{1}(V_1, E_1)$ is an induced subgraph of $G_{2}(V_2, E_2)$ and $V_2 = V_1 \ \bigcup \{v_0\}$, where $v_0$ is a vertex that is not in $V_1$. The following set of measurements, ${\cal M}_2 = \{\overline{M}_i\}_{i \in V_2}$, has $G_{2}(V_2, E_2)$ as its compatibility graph,
\begin{equation}
\overline{M}_i = M_i \bigotimes_{j\in V_1} \Pi_{i,j},\,\forall i \in V_1\;\;\text{and}\;\;\overline{M}_{v_0} = \mathbb{I}_d \bigotimes_{j\in V_1} P_{j},
\end{equation}
where
\begin{equation}
\Pi_{i,j} = \begin{cases}
\mathbb{I}_2 , & j\neq i,\\
|0\rangle\langle 0|, & j=i,
\end{cases}
\end{equation}
\begin{equation}
P_j = \begin{cases}
\mathbb{I}_2, & (v_0,j) \in E_2,\\
|\psi\rangle\langle \psi|, & (v_0,j)\not\in E_2,
\end{cases},
\end{equation}
$d$ is the dimension of each of the elements in $\{M_i\}_{i \in V_1}$, $\mathbb{I}_k$ is the identity operator in dimension $k$,
and $|\psi\rangle = (|0\rangle + |1\rangle)/\sqrt{2}$.
By construction, the state $\overline{\rho} = \rho \otimes_{j\in V_1} |0\rangle\langle 0|$ and the measurements $\overline{\mathcal{M}}_1 = \{\overline{M}_i\}_{i\in V_1}$ produce the same probabilities as the ones produced by $\rho$ and $\mathcal{M}_1$. That is, for every outcome $\overline{m_i}$ of $\overline{M}_i \in \overline{\mathcal{M}}_1$ and every outcome $m_i$ of $M_i \in \mathcal{M}_1$,
\begin{equation}
P( \overline{m}_i | \overline{\rho} ) = P ( m_i | \rho).
\end{equation}

This implies that, if there is an induced subgraph of a given compatibility graph $G$ that can produce contextuality, then the graph $G$ can produce at least the same amount of contextuality. Now, notice that a compatibility graph $G$ is not chordal if and only if it has induced cycles of size $k \ge 4$. Let us suppose that $C_k$ is one of them. If we could find a set of measurements ${\cal M}_C = \{M_i\}_{i \in V_C}$ whose compatibility graph were isomorphic to $C_k$ and that would produce contextuality, then, by the previous result, $G$ would also produce contextuality (at least the same amount as the induced $C_k$), thus proving our claim. For any $k$, explicit examples of sets of measurements satisfying all these requirements can be found in Ref.~\cite{AQB13}. \hfill \endproof

\begin{figure*}
	\hspace{-4mm}
	\includegraphics[width=0.93\linewidth]{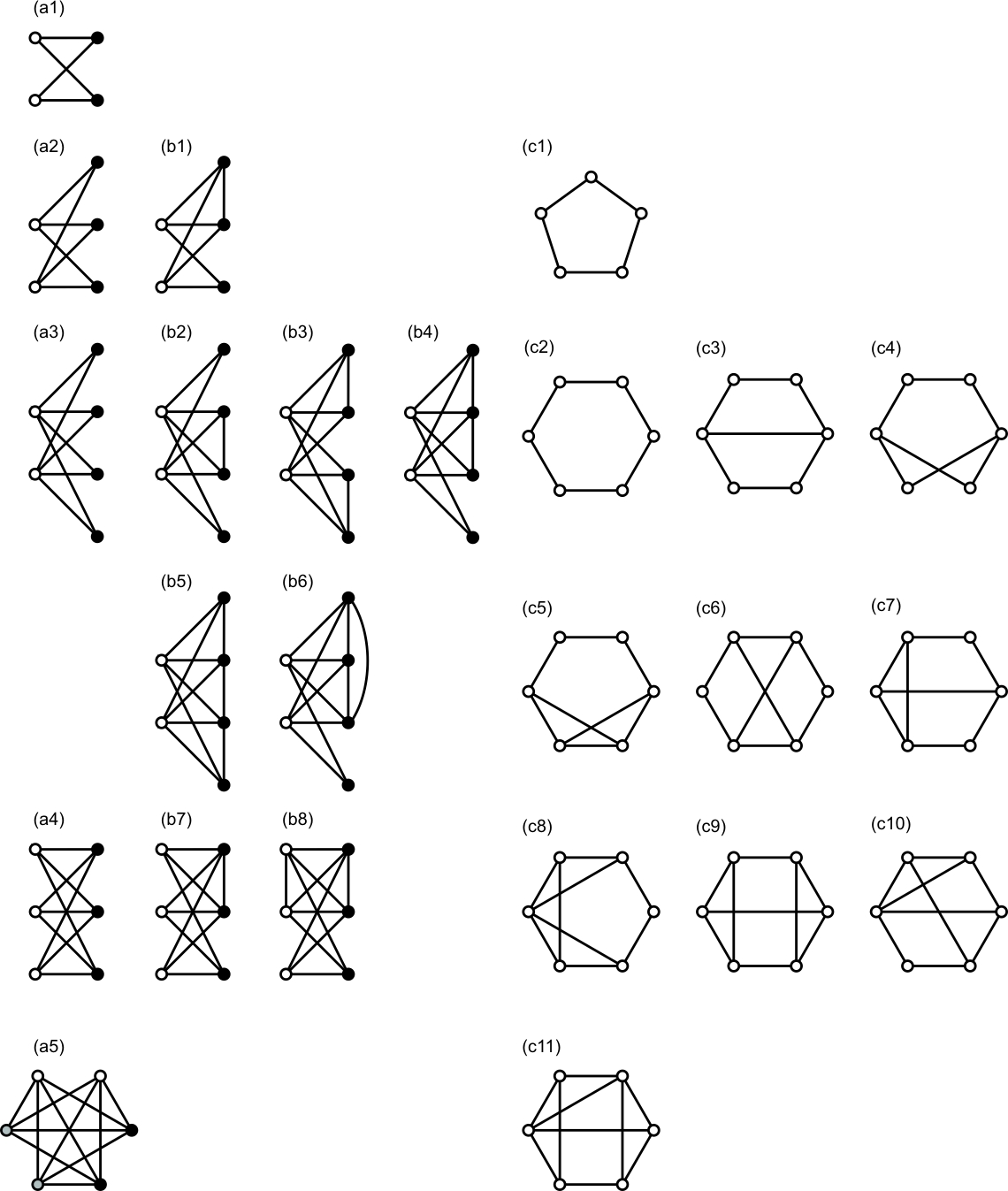}
	\caption{All the compatibility graphs corresponding to scenarios that can produce quantum contextuality with up to six ideal measurements. They are of three types: (a1)--(a5) can be realized in Bell scenarios (dots of the same color denote measurements performed by the same party); (b1)--(b8) can be realized in multipartite scenarios, but at the cost that at least one party has at least two compatible measurements, and (c1)--(c11) cannot be realized in multipartite scenarios.
	}
	\label{Fig1}
\end{figure*}

{\em Classification of the scenarios with quantum contextuality.} An interesting consequence of the previous theorem is that it allows us to identify and classify all measurement scenarios in which incompatibility can produce contextuality and tells us how to use quantum theory to produce contextuality in each of them. Given a fixed number $k$ of ideal measurements, to identify all scenarios that can produce contextuality, it is enough to compute all nonchordal graphs with $k$ vertices and avoid the cases in which one of the measurements is not needed for contextuality by removing those graphs in which one of the vertices does not belong to any cycle of length four or more. For $k$ up to $6$, the complete list of compatibility graphs corresponding to scenarios in which contextuality can occur is shown in Fig.~\ref{Fig1}.
All these compatibility graphs can be realized in experiments with sequential measurements on single systems, such as the experiments of Refs.~\cite{MWZ00,YLB03,KZG09,LMZNACH17}. In addition, some of the compatibility graphs can be realized in multipartite scenarios, since their sets of vertices can be divided into disjoint subsets, each subset corresponding to the measurements of one party and containing some nonadjacent vertices (i.e., incompatible measurements), and such that each vertex in a subset is adjacent to all vertices in the other subsets. According to this criterion, the graphs of compatibility that can produce quantum contextuality can be classified in three types:

(a) Nonchordal compatibility graphs that are complete $n$-partite, with $n\ge2$ (i.e., whose sets of vertices can be divided into $n$ disjoint and independent subsets such that each vertex in a set is adjacent to all vertices in the other subsets), as the graphs in Figs.~\ref{Fig1}(a1)--\ref{Fig1}(a5). If $n=2$, then the graphs have realizations as bipartite Bell scenarios. For example, the graphs in Figs.~\ref{Fig1}(a1)--\ref{Fig1}(a4). If $n=3$, then the graphs have realizations as tripartite Bell scenarios. For example, the graph in Fig.~\ref{Fig1}(a5). The sets of classical and quantum behaviors for these scenarios have been studied extensively, since the boundaries of the classical (noncontextual) sets are tight Bell inequalities. Specifically, for measurements with two outcomes, the exhaustive list of tight Bell inequalities that bound the set of classical behaviors for the scenario whose compatibility graph is in Fig.~\ref{Fig1}(a1) is in Refs.~\cite{CHSH69,Tsirelson80,SZ81,Froissart81,Fine82}, while the corresponding set of quantum behaviors is exhaustively characterized in Refs.~\cite{Tsirelson80,Tsirelson85,KC85,Tsirelson93,NPA07,NPA08}. Similarly, for the scenarios in Figs.~\ref{Fig1}(a2)--\ref{Fig1}(a3), the tight Bell inequalities and their quantum violations are presented in Ref.~\cite{BG08}, and in Ref.~\cite{CG04} for the scenario in Fig.~\ref{Fig1}(a4). Finally, the full set of tight Bell inequalities for the scenario in Fig.~\ref{Fig1}(a5) is in Refs.~\cite{PS01,Sliwa03} and the corresponding quantum violations in Ref.~\cite{LXC16}.

(b) Nonchordal compatibility graphs that have realizations as multipartite scenarios (since their vertices can be divided into disjoint sets, each of them containing some nonadjacent vertices, and such that each vertex in a subset is adjacent to all vertices in the other subsets), but in which at least one party has at least two measurements that are compatible (i.e., at least one of the subsets is not an independent set). These graphs are shown in Figs.~\ref{Fig1}(b1)--\ref{Fig1}(b8). So far, to our knowledge, these types of compatibility graphs have been considered only in relation with scenarios of nonlocality via local contextuality \cite{Cabello10,LHC16} and monogamy between nonlocality and local contextuality \cite{KCK14,ZZL16}. However, unlike in all these cases, in the scenarios in Figs.~\ref{Fig1}(b1)--\ref{Fig1}(b6), none of the parties has a set of measurements capable to locally produce contextuality, thus our result reveals a different form of quantum contextuality that is worth closer examination. Specifically, it would be interesting to compare the classical and quantum sets of behaviors with those of the scenario in Fig.~\ref{Fig1}(a1), since it seems that there are quantum behaviors that are contextual in the scenarios of Figs.~\ref{Fig1}(b1)--\ref{Fig1}(b6) but that are noncontextual when we ignore the measurements in Bob's side that are not in the scenario of Fig.~\ref{Fig1}(a1). However, this contextuality is not merely local (as occurs in Refs.~\cite{Cabello10,LHC16,KCK14,ZZL16}).

(c) Nonchordal compatibility graphs that do not admit realizations as multipartite scenarios (since their sets of vertices cannot be divided into disjoint subsets containing some nonadjacent vertices and such that each vertex in a subset is adjacent to all vertices in the other subsets). These graphs are shown in Figs.~\ref{Fig1}(c1)--\ref{Fig1}(c11). The most famous of them is the pentagon of compatibility shown in Fig.~\ref{Fig1}(c1), which corresponds to the scenario studied by Klyachko, Can, Binicio\u{g}lu, and Shumovsky (KCBS) \cite{KCBS08}. To our knowledge, so far, the classical and quantum sets of behaviors have been exhaustively characterized only for the scenarios corresponding to this graph and the hexagon in Fig.~\ref{Fig1}(c2) \cite{AQB13}. Our result allows us to identify new simple scenarios that can produce quantum contextuality. Curiously, the first Bell inequality different than the Clauser-Horne-Shimony-Holt inequality proposed in the literature, the two-party three-setting chained Bell inequality, proposed in Ref.~\cite{Pearle70} and rediscovered in Ref.~\cite{BC90}, is a tight noncontextuality inequality for a scenario corresponding to the compatibility graph in Fig.~\ref{Fig1}(c2) \cite{AQB13}, but is not a tight Bell inequality for the two-party three-setting Bell scenario corresponding to the compatibility graph in Fig.~\ref{Fig1}(a4).


{\em Conclusion.} Here, we have investigated the connection between the most basic form of nonclassicality---incompatibility---and the resource that has been proven to be necessary to explain the power of some leading models of quantum computation and many quantum information protocols---contextuality. We have proven that a necessary and sufficient condition for the existence of a quantum behavior that is contextual in the sense of Refs.~\cite{KCBS08,Cabello08,YO12,KBLGC12,AQB13,CSW14} is that the compatibility graph that encodes the relations of incompatibility between the measurements is nonchordal. Since being nonchordal implies containing induced cycles of size larger than three, our result points out the crucial role for quantum contextuality of the $n$-cycle compatibility scenarios with $n\ge 4$ (whose complete list of tight noncontextuality inequalities and their maximal quantum violation are presented in Ref.~\cite{AQB13}).

The scenarios in which contextuality can happen can be classified in three types: Bell scenarios, KCBS-type scenarios, and a third type in between them that worth closer examination.
This classification holds not only for quantum theory but for general probabilistic theories, as it is based on the observation that contextuality can only occur in scenarios whose compatibility graph is nonchordal, and nonlocality can only occur if, in addition, the vertices of the compatibility graph can be divided into disjoint sets, each of them containing only nonadjacent vertices. In fact, one of the interesting consequences of our result is the observation that what is special about quantum theory is that contextuality and nonlocality occur in all scenarios in which they can, respectively, occur.


\begin{acknowledgments}
	We thank Teiko Heinosaari, Andrei Khrennikov, and an anonymous referee of another paper for repeatedly asking the question addressed in this Rapid Communication, Costantino Budroni for bringing to our attention Vorob'yev's theorem, and Antonio J.\ L\'opez-Tarrida for comments.
	This work was supported by Project No.\ FIS2017-89609-P, ``Quantum Tools for Information, Computation and Research'' (MINECO-MICINN, Spain) with FEDER funds,
	the FQXi Large Grant ``The Observer Observed: A Bayesian Route to the Reconstruction of Quantum Theory,''
	and the project ``Photonic Quantum Information'' (Knut and Alice Wallenberg
	Foundation, Sweden). Z.-P.X.\ is supported by the Natural Science Foundation of China (Grant No.\ 11475089) and the China Scholarship Council.
\end{acknowledgments}



\appendix*


\section{Vorob'yev's theorem}


Here, we restate in the language of graph theory and prove a theorem introduced, without a proof, by Vorob'yev in 1963 \cite{Vorob'yev63} and then proven independently by Kellerer \cite{Kellerer64a,Kellerer64b}, Vorob'yev \cite{Vorob'yev67}, and others \cite{Malvestuto88}. Vorob'yev's theorem is the basis of a fundamental result in the field of expert systems \cite{LS88}.

Recall that a perfect elimination ordering in a graph $G$ is an ordering of the vertices of $G$ such that, for each vertex $v_i$, $v_i$ and the vertices of $G$ that are adjacent to $v_i$ and occur after $v_i$ in the order form a clique.
A graph is chordal if and only if it has a perfect elimination ordering \cite{FG65}.


{\em Theorem.}
	Any set of probabilities for the outcomes of a set of measurements whose compatibility relations are represented by a chordal graph admits a global extension to a joint probability distribution.


{\em Proof.}
	Suppose an $n$-vertex chordal graph $G$. Since $G$ is chordal, $G$ has a perfect elimination order $(v_n,v_{n-1},\ldots,v_1)$. Let $A_k$ be the set of vertices of $G$ that are adjacent to $v_k$ and occur after $v_k$ in that order. By definition of perfect elimination ordering, $A_k$ is a clique. Therefore, $\{M_v\}_{v\in A_k}$ is a set of mutually compatible measurements.
	Let $\{M_{v_i} = m_{v_i}\}_{i=1}^k$, for $i=1,2,\ldots,k$, be the set of events in which the output of measurement $M_{v_i}$ is $m_{v_i}$. Let us define
	\begin{widetext}
		\begin{eqnarray}
		P_1(M_{v_1} = m_{v_1}) &:= & \operatorname{Prob}(M_{v_1} = m_{v_1}),\\
		P_k(\{M_{v_i} = m_{v_i}\}_{i=1}^k) & := &
		\frac{\operatorname{Prob}(\{M_v = m_v\}_{v\in A_k})P_{k-1}(\{M_{v_i} = m_{v_i}\}_{i=1}^{k-1})}{\operatorname{Prob}(\{M_v = m_v\}_{v \in A_k\backslash v_k})},
		\end{eqnarray}
	\end{widetext}
	where $\operatorname{Prob}(\{M_v = m_v\}_{v\in A})$ denotes the probability distribution for a set of compatible measurements $\{M_v\}_{v\in A}$.

	We have to prove that $P_n(\{M_{v_i} = m_{v_i}\}_{i=1}^n)$ is a joint probability distribution which coincides with any $\operatorname{Prob}(\{M_v = m_v\}_{v\in A})$, where $A$ is a clique in $G$. We will prove it by induction.
	By definition, $P_1(M_{v_1} = m_{v_1})$ coincides with any $\operatorname{Prob}(\{M_v = m_v\}_{v\in A})$. Let us assume that $P_{t}(\{M_{v_i} = m_{v_i}\}_{i=1}^{t})$ also coincides with $\operatorname{Prob}(\{M_v = m_v\}_{v\in A})$, for $1 \leq t \leq k-1$ and any clique $A$ in $G$, that is,
	\begin{widetext}
		\begin{eqnarray}
		\sum_{m_v, v\in \{v_1,\ldots,v_{t}\}\backslash A} P_{t}(\{M_{v_i} = m_{v_i}\}_{i=1}^{t}) & = & \sum_{m_v, v\in A\backslash \{v_1,\ldots,v_{t}\}} \operatorname{Prob}(\{M_v = m_v\}_{v\in A}).
		\end{eqnarray}
	\end{widetext}
	Then, for any clique which does not contain $v_k$,
	\begin{widetext}
		\begin{eqnarray}
		\sum_{m_v, v\in \{v_1,\ldots,v_k\}\backslash A} P_k(\{M_{v_i} = m_{v_i}\}_{i=1}^k) &=& \sum_{m_v, v\in \{v_1,\ldots,v_{k-1}\}\backslash A} \frac{\left(\sum_{m_{v_k}}\operatorname{Prob}(\{M_v = m_v\}_{v\in A_k})\right)P_{k-1}(\{M_{v_i} = m_{v_i}\}_{i=1}^{k-1})}{\operatorname{Prob}(\{M_v = m_v\}_{v \in A_k\backslash v_k})}\\
		&=& \sum_{m_v, v\in \{v_1,\ldots,v_{k-1}\}\backslash A} P_{k-1}(\{M_{v_i} = m_{v_i}\}_{i=1}^{k-1})\\
		&=& \sum_{m_v, v\in A\backslash \{v_1,\ldots,v_{k-1}\}} \operatorname{Prob}(\{M_v = m_v\}_{v\in A})\\
		&=& \sum_{m_v, v\in A\backslash \{v_1,\ldots,v_k\}} \operatorname{Prob}(\{M_v = m_v\}_{v\in A}).
		\end{eqnarray}
	\end{widetext}
	If $A$ contains $v_k$, then $A \cap \{v_1,\ldots,v_k\} \subseteq A_k$ by definition. Therefore,
	\begin{widetext}
	\begin{eqnarray}
		\sum_{m_v, v\in \{v_1,\ldots,v_k\}\backslash A} P_k(\{M_{v_i} = m_{v_i}\}_{i=1}^k) &=& \sum_{m_v, v\in A_k\backslash A} \frac{\operatorname{Prob}(\{M_v = m_v\}_{v\in A_k})
			\left(\sum_{m_v, v\in \{v_1,\ldots,v_k\}\backslash A_k}P_{k-1}(\{M_{v_i} = m_{v_i}\}_{i=1}^{k-1})\right)}{\operatorname{Prob}(\{M_v = m_v\}_{v \in A_k\backslash v_k})}\\
		&=& \sum_{m_v, v\in A_k\backslash A} \operatorname{Prob}(\{M_v = m_v\}_{v\in A_k})\\
		&=& \sum_{m_v, v\in A\backslash \{v_1,\ldots,v_k\}} \operatorname{Prob}(\{M_v = m_v\}_{v\in A}).
	\end{eqnarray}
	\end{widetext}
	So $P_k(\{M_{v_i} = m_{v_i}\}_{i=1}^k)$ also coincides with any $\operatorname{Prob}(\{M_v = m_v\}_{v\in A})$, where $A$ is a clique in $G$. By induction, so does $P_n(\{M_{v_i} = m_{v_i}\}_{i=1}^n)$.\hfill \endproof


\end{document}